\documentclass[fleqn,twoside]{article}
\usepackage{espcrc2}
\usepackage{bm}
\usepackage{graphicx}
\usepackage{graphics}
\usepackage[figuresright]{rotating}

\newcommand{\AmS}{{\protect\the\textfont2
  A\kern-.1667em\lower.5ex\hbox{M}\kern-.125emS}}

\title{Multipion decays of $\omega(782)$ and $\phi(1020)$. }

\author{N.N. Achasov\address[MCSD]{Laboratory of Theoretical Physics,
        S.L. Sobolev Institute for Mathematics,
        630090, Novosibirsk, Russian Federation}
        and A.~A. Kozhevnikov\addressmark}

\begin{document}

\begin{abstract}
Using the chiral model of pseudoscalar, vector, and axial vector
mesons based on the hidden local symmetry added with the terms
induced by the Wess-Zumino anomaly,  the results of calculations
of the branching fractions of the decays $\omega(782)$ and
$\phi(1020)$ mesons to the $2\pi^+2\pi^-\pi^0$, $\pi^+\pi^-3\pi^0$
multipion states are presented. \vspace{1pc}
\end{abstract}

% typeset front matter (including abstract)
\maketitle

\section{INTRODUCTION}

The theory aimed at describing low energy hadron processes should
be formulated in terms of effective colorless degrees of freedom.
They can be introduced on the basis of spontaneously broken
approximate chiral symmetry $SU(3)_L\times SU(3)_R$ which is the
symmetry of QCD Lagrangian
\begin{eqnarray}
{\cal L}_{\rm QCD}&=&-\frac{1}{4}\left(\partial_\mu
G^a_\nu-\partial_\nu G^a_\mu+gf_{abc}G^b_\mu G^c_\nu\right)^2
+\nonumber\\&&\sum_{q=u,d,s,c,b,t}\bar
q\times\nonumber\\&&\left[\gamma_\mu\left(i\partial_\mu-
g\frac{\lambda^a}{2}G_{a\mu}\right)- m_q\right]q,
\end{eqnarray}
($[\lambda^a,\lambda^b]=2if_{abc}\lambda^c$) relative independent
rotations of right and left fields of  approximately massless
$u,d,s$ quarks:
\begin{eqnarray}
q_L\equiv\frac{1+\gamma_5}{2}q&\rightarrow&
V_Lq_L, \nonumber\\
q_R\equiv\frac{1-\gamma_5}{2}q&\rightarrow& V_Rq_R,
\end{eqnarray}
where $V_{L,R}\in SU(3)_{L,R}$. The pattern of the spontaneous
breaking is $SU(3)_L\times SU(3)_R\Rightarrow SU(3)_{L+R}$.
According to the Goldstone theorem, spontaneous breaking of global
symmetry results in appearance of massless fields. In our case
they are light $J^P=0^-$ mesons $\pi^+$, $\pi^-$, $\pi^0$, $K^+$,
$K^0$, $K^-$, $\bar K^0$, $\eta$. The transformation law
$U\rightarrow V_LUV^\dagger_R$ where
$U=\exp\left(i{\bm\Phi}\sqrt{2}/f_\pi\right)$, and
\begin{eqnarray}{\bm\Phi}&=&\left(
\begin{array}{ccc}
\frac{\pi^0}{\sqrt{2}}+\frac{\eta}{\sqrt{6}}&\pi^+&K^+\\
\pi^-&-\frac{\pi^0}{\sqrt{2}}+\frac{\eta}{\sqrt{6}}&K^0\\ K^-&\bar
K^0&-\frac{2\eta}{\sqrt{6}}
\end{array}\right),\;
\end{eqnarray}
fixes the Lagrangian of interacting Goldstone mesons:
$$
{\cal L}_{\rm GB}=\frac{f^2_\pi}{4}\mbox{Sp}\left(\partial_\mu
U\partial_\mu U^\dagger\right)+\cdots.$$ Dots mean the terms with
higher derivatives. Upon adding the term $\propto
m^2_\pi\mbox{Sp}(U+U^\dagger)$ which explicitly breaks chiral
symmetry, Goldstone bosons become massive. The Wess-Zumino (WZ)
term $$\Gamma_{\rm
WZ}=-\frac{in_c}{240\pi^2}\int_{M_5}\mbox{Sp}\left(dUU^\dagger\right)^5
$$removes spurious selection rule forbidding processes  with odd
number of Goldstone mesons.

Pseudoscalar mesons are produced in $e^+e^-$ annihilation via
vector resonances, hence one should  include vector mesons in a
chiral invariant way. The problem of testing chiral models of the
vector meson interactions with  Goldstone bosons is acute because
in well studied decays $\rho\to2\pi$, $\omega\to3\pi$ the final
pions are not soft enough to rely on the lowest derivative tree
effective Lagrangian. The multiple pion decays are most promising
because pions are truly soft.

\section{THE MODEL}

\begin{figure}[htb]
\includegraphics[scale=0.45]{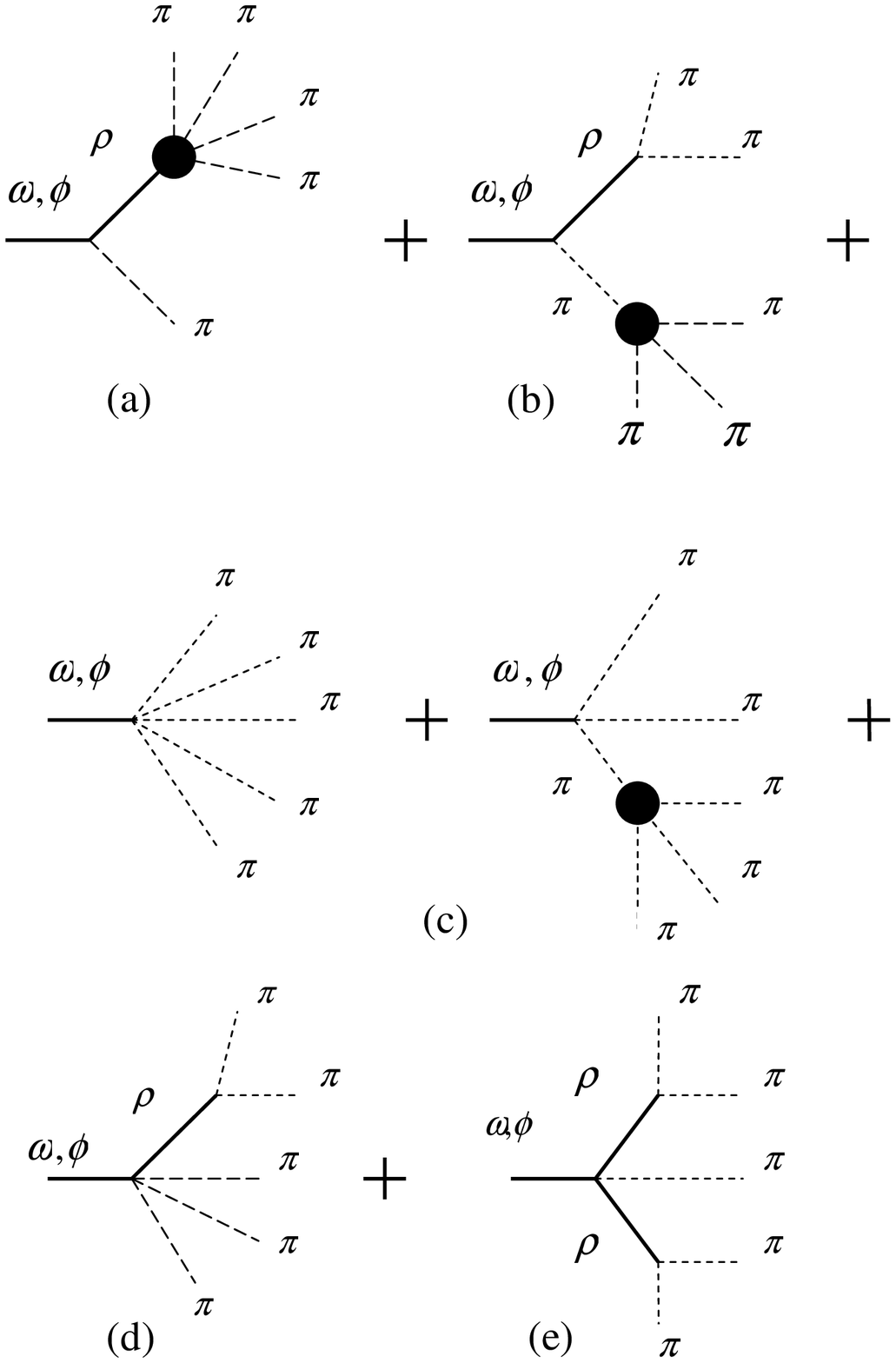}\caption{The diagrams
describing the $\omega,\phi\to5\pi$ decay. Shaded circles include
the vector meson exchange contributions \cite{ach03,ach05,ach00}.
All possible isotopic contributions are understood.}
\label{diagr5pi}
\end{figure}

There are a number of various chiral models incorporating
non-anomalous interactions of pseudoscalar, vector, and axial
vector mesons which are equivalent at the level of lowest number
of derivatives \cite{schechter,meissner88}. However, the anomalous
precesses are most conveniently treated in the framework of the
generalized hidden local symmetry (GHLS) approach  \cite{bando88}.
Choosing the specific gauge and restricting to the sector of
nonstrange mesons it looks as
\begin{eqnarray}
{\cal L}&=&a_0f^2_\pi{\rm
Tr}\left(\frac{\partial_\mu\xi^\dagger\xi+\partial_\mu\xi\xi^\dagger}{2i}-gV_\mu\right)^2+
\nonumber\\&&b_0f^2_\pi{\rm
Tr}\left(\frac{\partial_\mu\xi^\dagger\xi-\partial_\mu\xi\xi^\dagger}{2i}+gA_\mu\right)^2
+\nonumber\\ &&c_0f^2_\pi g^2{\rm Tr}A^2_\mu+d_0f^2_\pi{\rm
Tr}\left(\frac{\partial_\mu\xi^\dagger\xi-\partial_\mu\xi\xi^\dagger}{2i}\right)^2
\nonumber\\&&-\frac{1}{2}{\rm
Tr}\left(F^{(V)2}_{\mu\nu}+F^{(A)2}_{\mu\nu}\right)\nonumber\\&&
-i\alpha_4g{\rm
Tr}[A_\mu,A_\nu]F^{(V)}_{\mu\nu}+2i\alpha_5g\times\nonumber\\&&{\rm
Tr}\left(\left[\frac{\partial_\mu\xi^\dagger\xi-\partial\xi\xi^\dagger}{2ig},A_\nu\right]
+[A_\mu,A_\nu]\right)\times\nonumber\\&&F^{(V)}_{\mu\nu},
\end{eqnarray}
where $\xi=\exp i\frac{{\bm\tau}\cdot{\bm\pi}}{2f_\pi}$,
\begin{eqnarray}
F^{(V)}_{\mu\nu}&=&\partial_\mu V_\nu-\partial_\nu
V_\mu-i[V_\mu,V_\nu]-i[A_\mu,A_\nu],\nonumber\\
F^{(A)}_{\mu\nu}&=&\partial_\mu A_\nu-\partial_\nu
A_\mu-i[V_\mu,A_\nu]-i[A_\mu,V_\nu],\nonumber\\
V_\mu&=&\left(\frac{{\bm\tau}}{2}\cdot{\bm\rho}_\mu\right)+\frac{\omega_\mu}{2},\nonumber\\
A_\mu&=&\left(\frac{{\bm\tau}}{2}\cdot{\bm A }_\mu\right).
\end{eqnarray}
The following steps are necessary in order to obtain the minimal
effective Lagrangian describing non-anomalous processes:
\begin{itemize}
\item To exclude axial vector-pseudoscalar mixing by means of
introduction of physical $a_1(1260)$ meson field $a_\mu$
$$
A_\mu=a_\mu-\frac{b_0}{g(b_0+c_0)}\frac{\partial_\mu\xi^\dagger\xi-\partial_\mu\xi\xi^\dagger}
{2i}.$$ Here  the total nonlinear combination of pion fields is
rotated away. This provides the fulfillment the Adler condition
for  decay amplitudes with the intermediate $a_1$ meson
independently of its mass. \item To renormalize according to
$f_\pi\to Z^{-1/2}f_\pi$, ${\bm\pi}\to Z^{-1/2}{\bm\pi}$,
$(a_0,b_0,c_0,d_0)=Z(a,b,c,d),$ where
$$\left(d_0+\frac{b_0c_0}{b_0+c_0}\right)Z^{-1}=1.$$
\item To make the choice $\alpha_4=-\alpha_5=-1$, $a=b=c=2$, $d=0$
which removes the higher derivative $\rho\pi\pi$ coupling, results
in universality $g_{\rho\pi\pi}=g$, vector dominance, KSRF
$2g^2_{\rho\pi\pi}f^2_\pi/m^2_\rho=1$.
\end{itemize}
Anomalous vertices $\omega\to3\pi,5\pi,\rho\pi,\rho\rho\pi$ etc.
are treated  using the lagrangian induced by the Wess-Zumino
anomaly equation. We give the necessary lowest derivative terms of
effective lagrangian:
\begin{eqnarray}
\cal{L}^{\rm
an}&=&\frac{n_cg}{32\pi^2f^3_\pi}\varepsilon_{\mu\nu\lambda\sigma}\left\{\left[c_1-c_2-c_3
+\frac{{\bm\pi}^2}{4f^2_\pi}\times\right.\right.\nonumber\\&&\left.\left.
\left(\frac{5}{3}(c_2+c_3)-c_1\right)\right]
\omega_\mu\partial_\nu\bm{\pi}\cdot\right.\nonumber\\&&
\left.[\partial_\lambda\bm{\pi}\times\partial_\sigma\bm{\pi}]
-g(c_1+c_2-c_3)\omega_\mu\times\right.\nonumber\\&&\left.\left[({\bm\rho}_\lambda
\cdot\partial_\sigma{\bm\pi})({\bm\pi}\cdot\partial_\sigma{\bm\pi})-
gf^2_\pi([{\bm\rho}_\nu\times{\bm\rho}_\lambda]\cdot\right.\right.\nonumber\\&&
\left.\left.\partial_\sigma{\bm\pi})\right]
-4c_3gf^2_\pi\partial_\mu\omega_\nu\left[({\bm\rho}_\lambda\cdot\partial_\sigma{\bm\pi})+
\right.\right.\nonumber\\&&\left.\left.\frac{1}{6f^2_\pi}\left([{\bm\rho}_\lambda\times{\bm\pi}]
\cdot[{\bm\pi}\times\partial_\sigma{\bm\pi}]\right)
\right]\right\}.
\end{eqnarray}
The diagrams describing the $\omega\to5\pi$ decay amplitude are
shown in Fig.~\ref{diagr5pi}. The obtained decay amplitudes
satisfy the Adler condition. See Refs.~\cite{ach03,ach05,ach00}
for more detail.
%%%%%%%%%%%%%%%%%%%%%%%%%%%%%%%%%%%%%%%%%%%%%%%%%%%%%%%
%%%%%%%%%%%%%%%%%%%%%%%%%%%%%%%%%%%%%%%%%%%%%%%%%%%%%%%%%%%%%%%%%%

The parameters $c_{1,2,3}$ are arbitrary. They are fixed in accord
with  the condition of absence of the pointlike
$\omega\to\pi^+\pi^-\pi^0$ vertex, hence $c_1-c_2-c_3=0$. Then
$c_3$ is fixed by the  coupling constant
$g_{\omega\rho\pi}=-\frac{n_cg^2c_3}{8\pi^2f_\pi}$. Next we take
$c_1+c_2-c_3=0$, which  seems arbitrary, but within the accuracy
to percent the results are robust to wide variations  breaking
this condition \cite{ach03}. So, our choice is
\begin{equation}
c_1=c_3\mbox{, }c_2=0.\label{param}\end{equation}

\section{THE RESULTS}

The $\omega,\phi\to5\pi$ decay width is represented as
8-dimensional integral over Mandelstam-like Kumar variables:
\begin{eqnarray}
\Gamma_{5\pi}(s)&=&\frac{\pi^2\sqrt{s}}{24 N_{\rm
sym}\times(2\pi)^{11}}\int_{s_{1-}}^{s_{1+}}ds_1\times\nonumber\\&&\int_{s_{2-}}^{s_{2+}}ds_2
\int_{s_{3-}}^{s_{3+}}ds_3\int_{u_{1-}}^{u_{1+}}du_1\times\nonumber\\&&\int_{u_{2-}}^{u_{2+}}
\frac{du_2}{[\lambda(s,s_2,s^\prime_2)\lambda(s,m^2_3,u_2)]^{1/2}}
\times\nonumber\\&&\int_{u_{3-}}^{u_{3+}}
\frac{du_3}{[\lambda(s,s_3,s^\prime_3)\lambda(s,m^2_4,u_3)]^{1/2}}\times\nonumber
\\&&\int_{t_{2-}}^{t_{2+}}
dt_2[\lambda(s,t_1,t^\prime_1)(1-\xi^2_2)(1-\eta^2_2)\times\nonumber\\&&(1-\zeta^2_2)]^{-1/2}
\int_{t_{3-}}^{t_{3+}}dt_3|M_{\omega,\phi\to5\pi}|^2\times\nonumber
\\&&
[\lambda(s,t_2,t^\prime_2)(1-\xi^2_3)(1-\eta^2_3)\times\nonumber
\\&&
(1-\zeta^2_3)]^{-1/2}. \end{eqnarray} The notations are given in
Ref.~\cite{kumar}. The results of the evaluation of the branching
fractions of the $\omega\to5\pi$ decay modes under assumption
(\ref{param}) are collected in Table \ref{omegaresults}.
%%%%%%%%%%%%%%%%%%%%%%%%%%%%%%%%%%%%%%%%%%%%%%%%%%%%%%%%%%%%%%%%%%%%%%%%
\begin{table}
\caption{\label{omegaresults}The $\omega\to5\pi$ branching
fractions evaluated for different masses of the $a_1$ meson.}
\begin{tabular}{ccc}\hline
$m_{a_1}$
[GeV]&$B_{\omega\to\pi^+\pi^-3\pi^0}$&$B_{\omega\to2\pi^+2\pi^-\pi^0}$\\
\hline
1.09&$4.2\times10^{-9}$&$3.8\times10^{-9}$\\
1.23&$4.1\times10^{-9}$&$3.7\times10^{-9}$\\no
$a_1$&$3.6\times10^{-9}$&$3.3\times10^{-9}$\\  \hline
\end{tabular}
\end{table}
%%%%%%%%%%%%%%%%%%%%%%%%%%%%%%%%%%%%%%%%%%%%%%%%%%%%%%%%%%%%%%%%%%%%%%%%%%

The case of the $\phi\to5\pi$ decay is more subtle. If the $\phi$
transitions to nonstrange mesons are  due to $\phi\omega$ mixing
then the results of the $\omega\to5\pi$ calculations  are
translated to the case of $\phi$:
$$\Gamma_{\phi\to5\pi}(m^2_\phi)=|\varepsilon_{\phi\omega}|^2
\Gamma_{\omega\to5\pi}(m^2_\phi),$$
$$|\varepsilon_{\phi\omega}|^2=
\frac{\Gamma_{\phi\to3\pi}(m^2_\phi)
W_{\omega\to\rho\pi\to3\pi}(m^2_\omega)}
{\Gamma_{\omega\to3\pi}(m^2_\omega)W_{\phi\to\rho\pi\to3\pi}(m^2_\phi)}
=3\times10^{-3},$$where $W_{\phi,\omega\to\rho\pi\to3\pi}$ is the
dynamical phase space volume.  In the opposite case when $\phi$
goes to nonstrange mesons directly \cite{ach95}, one could
construct the effective Lagrangian guided by the property of the
decay amplitude being chiral invariant, by analogy with the
$\omega\to5\pi$ case and with the same assumptions about free
parameters \cite{ach03}. The results of calculations are collected
in Table \ref{phiresults}.

Two remarks are in order. First, the KLOE data on $\phi\to3\pi$
\cite{kloe} allow possible nonzero pointlike vertex. Taking  this
vertex into account  results in $\pm8\%$ deviation from the
figures obtained under declared assumption about free parameters
of effective Lagrangian. Second, the sum of the non-resonant
diagrams Fig.~\ref{diagr5pi}(b), (c), (d), (e) contributes about
$20\%$ to the $\phi\to5\pi$ partial width. This sum is incoherent
with the contribution of the dominant diagram
Fig.~\ref{diagr5pi}(a). Hence, within the accuracy $\pm20\%$ the
dynamics of the decay $\phi\to5\pi$ is dominated by the process
$\phi\to\rho\pi$ followed by the decay $\rho\to4\pi$ \cite{ach00}
with the resonant $\rho$ and thus is insensitive neither to
arbitrary parameters nor to the $\phi\omega$ mixing model.
%%%%%%%%%%%%%%%%%%%%%%%%%%%%%%%%%%%%%%%%%%%%%%%%%%%%%%%%%%%%%
\begin{table}
\caption{\label{phiresults}The same as in Table
\ref{omegaresults}, but for $\phi\to5\pi$.}
\begin{tabular}{ccc}\hline
$m_{a_1}$
[GeV]&$B_{\phi\to\pi^+\pi^-3\pi^0}$&$B_{\phi\to2\pi^+2\pi^-\pi^0}$\\
\hline
1.09&$4.4\times10^{-7}$&$8.8\times10^{-7}$\\
1.23&$3.9\times10^{-7}$&$7.7\times10^{-7}$\\no
$a_1$&$2.5\times10^{-7}$&$5.0\times10^{-7}$\\  \hline
\end{tabular}
\end{table}
%%%%%%%%%%%%%%%%%%%%%%%%%%%%%%%%%%%%%%%%%%%%%%%%%%%%%%%%%%%%%%%

The mass spectra of the four pion subsystem in the final five-pion
state are shown in Fig.~\ref{spec5pic} and \ref{spec5pin}.  Both
spectra has the peak due to the $\rho$ pole. The spectrum of
$\pi^+\pi^-\pi^-\pi^0$ has  the second  peak due to the
combination of the  strong energy dependent anomaly induced
contribution $\rho^-\to\omega\pi^-\to\pi^+\pi^-\pi^-\pi^0$ in the
decay $\phi\to\rho^-\pi^+\to\pi^+\pi^+\pi^-\pi^-\pi^0$ and the
phase space kinematical restriction. Such contribution is absent
in the decay $\phi\to\rho^-\pi^+\to\pi^+\pi^-\pi^0\pi^0\pi^0$.
\begin{figure}[htb]
\includegraphics[scale=0.55]{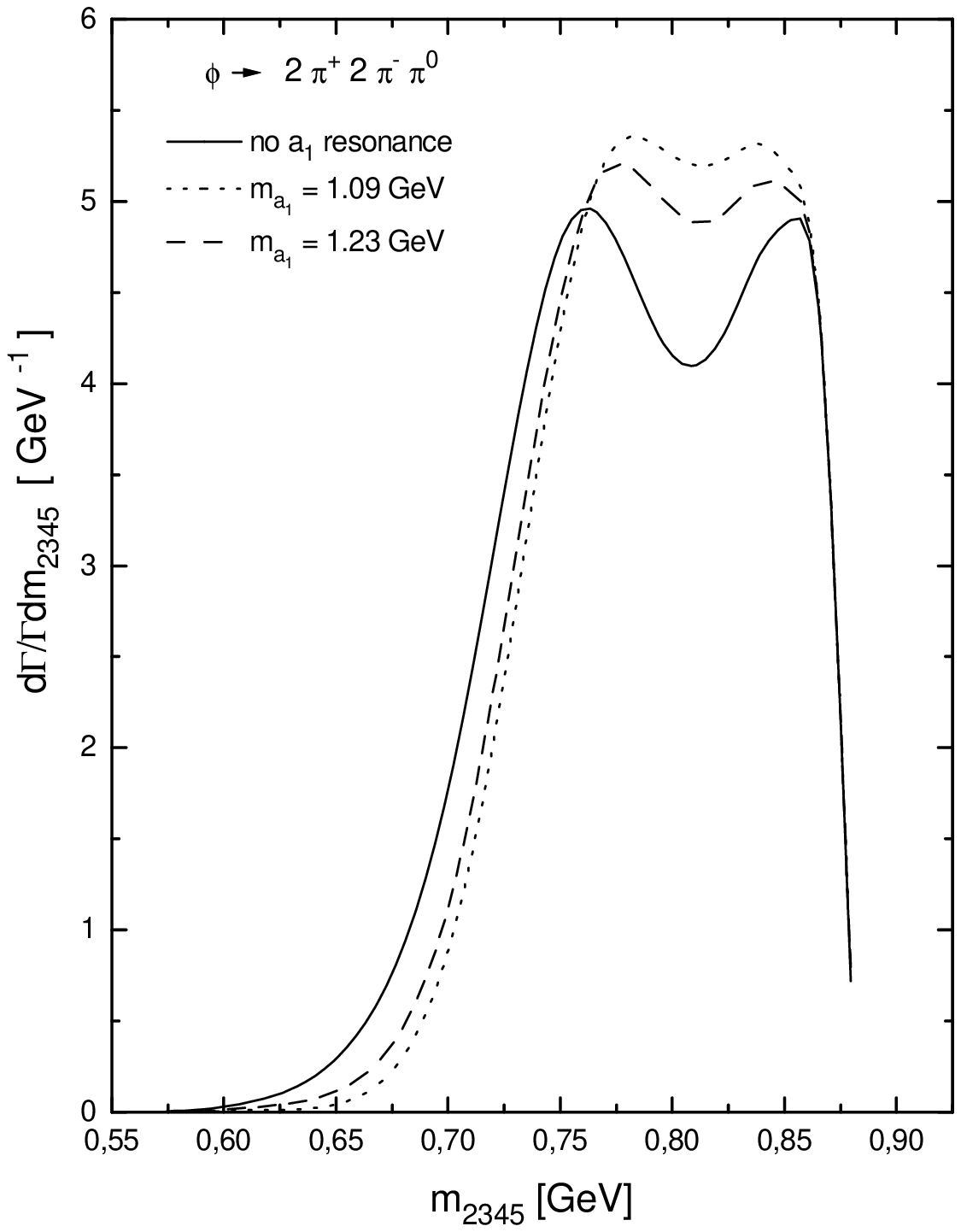}\caption{\label{spec5pic}
The mass spectrum of
$\pi^+_{q_2}\pi^-_{q_3}\pi^-_{q_4}\pi^0_{q_5}$ in the decay
$\phi\to\pi^+_{q_1}\pi^+_{q_2}\pi^-_{q_3}\pi^-_{q_4}\pi^0_{q_5}$
normalized to the respective $5\pi$  width; $\sqrt{s}=m_\phi$,
$m^2_{2345}=(q_2+q_3+q_4+q_5)^2\equiv s_1$.} \label{spec5pic}
\end{figure}
\begin{figure}[htb]
\includegraphics[scale=0.55]{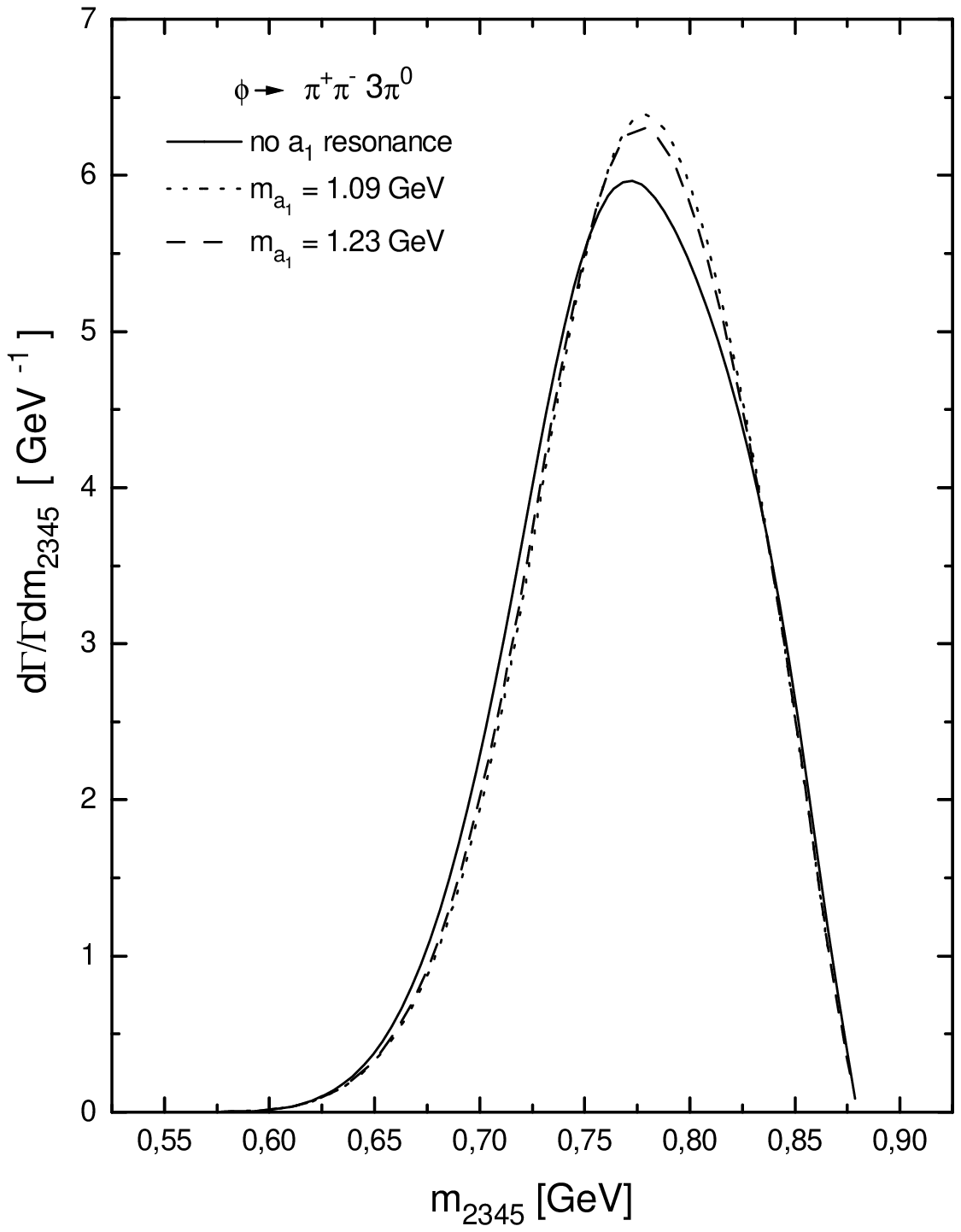}\caption{\label{spec5pin}
The same as in Fig.~\ref{spec5pic}, but for the subsystem
$\pi^-_{q_2}\pi^0_{q_3}\pi^0_{q_4}\pi^0_{q_5}$ in the decay
$\phi\to\pi^+_{q_1}\pi^-_{q_2}\pi^0_{q_3}\pi^0_{q_4}\pi^0_{q_5}$.}
\label{spec5pin}
\end{figure}
%%%%%%%%%%%%%%%%%%%%%%%%%%%%%%%%%%%%%%%%%%%%%%%%%%%%%%%%%%
With the total KLOE statistics  $\int{\cal L}dt\approx500$
pb$^{-1}$ (circa 2003) one can have the number of events of the
$\phi\to5\pi$ decay approximately  1340, 2070, 2360, respectively,
in the model without $a_1$, in the model incorporating $a_1$ with
the mass $m_{a_1}=1.23$ GeV, 1.09 GeV.
%%%%%%%%%%%%%%%%%%%%%%%%%%%%%%%%%%%%%%%%%%%%%%%%%%%%%%%%%%%%%

\end{document}